\documentclass[preprint,3p,12pt]{elsarticle}

\begin{document}

\title{\vskip-3cm{\baselineskip14pt
    \begin{flushleft}
      \normalsize SFB/CPP-12-03\\
      \normalsize TTP12-02
  \end{flushleft}}
  \vskip1.5cm
  {\tt CRunDec}: a {\tt C++} package for running and decoupling of the
  strong coupling and quark masses
}

\author{
  Barbara Schmidt 
  and
  Matthias Steinhauser
  \\[1em]
  {\small\it Institut f{\"u}r Theoretische Teilchenphysik}\\
  {\small\it Karlsruhe Institute of Technology (KIT)}\\
  {\small\it 76128 Karlsruhe, Germany}
}

\date{}

\begin{abstract}
In this paper we present the {\tt C++} package
{\tt CRunDec} which implements all relevant formulae needed
for the running and decoupling for the strong coupling constant and
light quark masses. Furthermore, several formulae are implemented which
can be used to transform the heavy quark masses among different
renormalization schemes.
{\tt CRunDec} is the {\tt C++} version on the {\tt Mathematica}
package {\tt RunDec} containing several updates and improvements.

\medskip


\end{abstract}

\maketitle


\newpage


\section*{Program summary}

\begin{itemize}

\item[]{\it Title of program:}
  {\tt CRunDec}

\item[]{\it Available from:}\\
  {\tt
  http://www-ttp.physik.uni-karlsruhe.de/Progdata/ttp12/ttp12-02/
  }

\item[]{\it Computer for which the program is designed and others on which it
    is operable:}
  Any computer where a {\tt C++} compiler is running.

\item[]{\it Operating system or monitor under which the program has been
    tested:} 
  Linux, Windows

\item[]{\it No. of bytes in distributed program including test data etc.:}
  $85\,000$

\item[]{\it Distribution format:} 
  source code

\item[]{\it Keywords:} 
  Quantum Chromodynamics, running coupling constant,
  running quark mass, on-shell mass, $\overline{\rm MS}$ mass,
  decoupling of heavy particles

\item[]{\it Nature of physical problem:}
  The value for the coupling constant of Quantum Chromodynamics,
  $\alpha_s^{(n_f)}(\mu)$,
  depends on the considered energy scale, $\mu$, and the number of active
  quark flavours, $n_f$. The same applies to light quark masses,
  $m_q^{(n_f)}(\mu)$,
  if they are, e.g., evaluated in the $\overline{\rm MS}$ scheme.
  In the program {\tt CRunDec} all 
  relevant formulae are collected and various procedures are provided
  which allow for a convenient evaluation of $\alpha_s^{(n_f)}(\mu)$
  and $m_q^{(n_f)}(\mu)$ using the state-of-the-art correction terms.

\item[]{\it Method of solution:}
  {\tt CRunDec} is implemented in {\tt C++}. For the solution of the 
  differential equations an adaptive Runge-Kutta procedure has been
  implemented.
  
\item[]{\it Restrictions on the complexity of the problem:}
  It could be that for an unphysical choice of the input parameters the 
  results are nonsensical.

\item[]{\it Typical running time:}
  In general the run time for the individual operations is below a
  millisecond. 

\end{itemize}


\section{Introduction}

Among the fundamental quantities of Quantum Chromodynamics there are
certainly the anomalous dimensions which control the scale dependence
of the parameters and fields. In this context a particular role is
taken over by the beta function and anomalous mass dimension in the
$\overline{\rm MS}$ scheme which govern the running of the strong
coupling constant $\alpha_s$ and the quark masses $m_q$.
Every time a flavour threshold is crossed in the running process 
decoupling relations have to be applied which guarantee that the heavy
quark is integrated out from the theory.
More than ten years ago the {\tt Mathematica} package {\tt RunDec} has
been written~\cite{Chetyrkin:2000yt} which incorporates all relevant formulae.
However, for many application it is more convenient to have the running
and decoupling routines available in the framework of 
a commonly used programming language and not within a computer
algebra system. Thus we have decided to convert the most important routines
to {\tt C++} adding at the same time new routines and improving others
w.r.t. the {\tt Mathematica} version. 
In what follows we describe the {\tt C++} file {\tt CRunDec} which
constitutes a {\tt C++} class and contains the functions known from
{\tt RunDec} as public components. 

The use of {\tt CRunDec} does not require any knowledge about object
oriented programming. The following
skeletons exemplify the usage and can easily be adapted to the problem at
hand. It is possible to work with pointers to an object of the type
{\tt CRunDec} and access the member functions correspondingly:
\begin{verbatim}
#include <iostream>
#include "CRunDec.h"
using namespace std;
int main(){
  CRunDec * <pointer> = new CRunDec();
  double <result> = <pointer> -> <function>(<parameters>);
  return(0);
}
\end{verbatim}
Alternatively also the following realization is possible:
\begin{verbatim}
#include <iostream>
#include "CRunDec.h"
using namespace std;
int main(){
  CRunDec <object>;
  double <result> = <object>.<function>(<parameters>);
  return(0);
}
\end{verbatim}
Explicit examples will be given below.

The remainder of the paper is organized as follows: In the next
Section all available functions are described,
Section~\ref{sec::examples} contains typical examples which exemplify
the usage of {\tt CRunDec}.


\section{Structure of {\tt CRunDec}}

All public components of the {\tt C++} class 
{\tt CRunDec} consist of functions which have
the same name as the corresponding function in the 
{\tt Mathematica} version~\cite{Chetyrkin:2000yt}.
In the following we list the function header (including the variable
names; see also Tab.~\ref{tab::var}) 
which --- together with the description in the Appendix of
Ref.~\cite{Chetyrkin:2000yt} --- specifies both the
usage and the purpose of the function.
There are some additions implemented in {\tt CRunDec} which are also 
described.

Let us mention that all functions listed in
Subsections~\ref{sub::run}--\ref{sub::dec} are overloaded w.r.t. the
argument $n_f$ (number of active flavours). I.e., this argument can be
omitted in case $n_f$ is specified  in the declaration of the {\tt
  CRunDec} object (see also examples in Section~\ref{sec::examples}).
In this context two auxiliary functions are quite useful: 
\verb|GetNf()| returns the specified number of active
flavours and \verb|SetNf(int nf)| can be used to set the number of
active flavours.

\begin{table}[tb]
  \begin{center}
    \begin{tabular}{c|c|l}
      \hline
      symbol in {\tt C++} code & mathematical symbol & meaning \\
      \hline
      \verb|asmu, asmu0, asmu1| & $\alpha_s(\mu)$, $\alpha_s(\mu_0)$, 
      $\alpha_s(\mu_1)$ 
      & strong coupling constant
      \\
      \verb|mq, mq0|     & $m_q(\mu)$, $m_q(\mu_0)$ & $\overline{\rm MS}$ quark mass
      \\
      \verb|mOS, mMS| & $M_q$, $m_q$ & on-shell and $\overline{\rm MS}$ quark mass
      \\
      \verb|mRI| & $m^{\rm RI}$ & regularization invariant mass
      \\
      \verb|mRGI| & $\hat{m}$ & renormalization group invariant mass
      \\
      \verb|Mth| & $M_{\rm th}$ & heavy quark mass
      \\
      \verb|mu, mu0, mu1|     & $\mu$, $\mu_0$, $\mu_1$ 
      & renormalization scale
      \\
      \verb|muth|     & $\mu_{\rm th}$
      & decoupling scale
      \\
      \verb|nf| & $n_f$ & number of active flavours
      \\
      \verb|nloops| & --- & number of loops
      \\
      \hline
    \end{tabular}
    \caption{\label{tab::var}Meaning of the variables used in the
      function headers.}
  \end{center}
\end{table}

\subsection{Input parameters}

For convenience of the user some frequently used input parameters 
are pre-defined in the file {\tt CRunDec.h} and can be used during the
calculation. They
read~\cite{Chetyrkin:2009fv,Nakamura:2010zzi,Lancaster:2011wr,Bethke:2011tr}
\begin{verbatim}
#define asMz 0.1183
#define Mz   91.18
#define Mt   173.2
#define Mb   4.8
#define Mc   1.5
#define muc  1.279
#define mub  4.163
#define Mtau 1.777
\end{verbatim}
If not stated otherwise these numbers are used in the examples presented 
in Section~\ref{sec::examples}.
In case other numerical values shall be used it is straightforward to
redefine the preprocessor variables.
The examples given in the appendix of Ref.~\cite{Chetyrkin:2000yt}
can be reproduced with
\begin{verbatim}
#define asMz 0.118
#define Mz   91.18
#define Mt   175.
#define Mb   4.7
#define Mc   1.6
#define muc  1.2
#define mub  3.97
#define Mtau 1.777
\end{verbatim}

\subsection{\label{sub::run}Functions related to the running of $\alpha_s$ and $m_q$}

\begin{itemize}
\item \verb| double LamExpl(double asmu, double mu, int nf, int nloops);|
\item \verb| double LamImpl(double asmu, double mu, int nf, int nloops);|
\item \verb| double AlphasLam(double Lambda, double mu, int nf, int nloops);|
\item \verb| double AlphasExact(double asmu0, double mu0, double mu1,|
  \\
      \verb|                    int nf, int nloops);|
\item \verb| double mMS2mMS(double mq0, double asmu0, double asmu1,| 
  \\
      \verb|                int nf, int nloops);|
\item \verb| AsmMS AsmMSrunexact(double mq0, double asmu0, double mu0,|
  \\
      \verb|                     double mu, int nf, int nloops);|
\end{itemize}

The function \verb|AsmMSrunexact| is new in {\tt CRunDec}. It solves
simultaneously the differential equations for $\alpha_s$ and $m_q$
with initial values $m_q(\mu_0)$ and $\alpha_s(\mu_0)$ and $n_f$ active
quark flavours. The return type of \verb|AsmMSrunexact| is a struct
with two double components, $\alpha_s(\mu)$ and $m_q(\mu)$. The
corresponding code in {\tt CRunDec} looks as follows: 
\begin{verbatim}
struct AsmMS {
      double Asexact;
      double mMSexact; 
};
\end{verbatim}
For convenience of the user there is a pre-defined variable
\verb|AsmMS AM|. Both components of \verb|AM| are initialized to zero when
creating a \verb|CRunDec| object.

\subsection{Functions relating different mass definitions}

\begin{itemize}
\item
  \verb|double mOS2mMS(double mOS, double mq[], double asmu, double mu,|
  \\
  \verb|               int nf, int nloops);|
\item
  \verb|double mMS2mOS(double mMS, double mq[], double asmu, double mu,|
  \\
  \verb|               int nf, int nloops);|
\item
  \verb|double mOS2mMSrun(double mOS, double mq[], double asmu, double mu,|
  \\
  \verb|                  int nf,int nloops);|
\item
  \verb|double mMS2mOSrun(double mMS, double mq[], double asmu, double mu,|
  \\
  \verb|                  int nf,int nloops);|
\item
  \verb|double mOS2mMSit(double mOS, double mq[], double asmu, double mu,|
  \\
  \verb|                 int nf,int nloops);|
\item
  \verb|double mOS2mSI(double mOS, double mq[], double asM,|
  \\
  \verb|               int nf, int nloops);|
\item
  \verb|double mMS2mSI(double mMS, double asmu, double mu,|
  \\
  \verb|              int nf, int nloops);|
\item
  \verb|double mMS2mRI(double mMS, double asmu, int nf, int nloops);|
\item
  \verb|double mRI2mMS(double mRI, double asmu, int nf, int nloops);|\footnote{Note
    a typo in the example to {\tt mRI2mMS} in the Appendix of
    Ref.~\cite{Chetyrkin:2000yt}:
    {\tt mRI2mMS[175,0.107,175,6,3]} should read 
    {\tt mRI2mMS[175,0.107,6,3]}.}
\item
  \verb|double mMS2mRGI(double mMS, double asmu, int nf, int nloops);|
\item
  \verb|double mRGI2mMS(double mRGI, double asmu, int nf, int nloops);|
\item
  \verb|double mMS2mRGImod(double mMS, double asmu, int nf, int nloops);|
\end{itemize}

Note that the light-quark-mass effects can be taken into account with
the help of the array \verb|mq[]| which is defined as
\begin{verbatim}
double mq[4];
\end{verbatim}
By default all elements of \verb|mq[]| are zero. In case non-zero
values are needed the array has to be filled before
the corresponding function is called.

In {\tt CRunDec} the implementation of \verb|mMS2mSI| has been
modified as compared to the {\tt Mathematica} version. It is now based
on \verb|AsmMSrunexact| and avoids the computation of $\Lambda_{\rm
QCD}$ in intermediate steps 
which is perturbatively more stable, in particular for lower
renormalization scales. Similar modifications have been performed
in \verb|mMS2mOSrun|.

The function \verb|mMS2mRGImod| is new in {\tt CRunDec}. It is defined
in analogy to \verb|mMS2mRGI|, however, the more commonly used 
convention has been adopted where the function $c(x)$ in Eq.~(11) of
Ref.~\cite{Chetyrkin:2000yt} is evaluated for $x=2\beta_0\alpha_s/\pi$ instead
of $x=\alpha_s/\pi$.

\subsection{\label{sub::dec}Functions related to the decoupling of heavy quarks}

As compared to the {\tt Mathematica} version {\tt CRunDec}
contains the decoupling relations only for the case of on-shell heavy
quarks which are most relevant for the practical purposes.
Furthermore, the functions \verb|DecLambdaUp| and 
\verb|DecLambdaDown| have not been implemented in the {\tt C++}
version since it is recommended to use \verb|AlL2AlH| and
\verb|AlH2AlL| in case a flavour threshold is crossed during the
running of $\alpha_s$.

\begin{itemize}
\item
  \verb|double DecAsDownOS(double asmu, double Mth, double muth,|
  \\
  \verb|                   int nf, int nloops);|
\item 
  \verb|double DecAsUpOS(double asmu, double Mth, double muth,|
  \\
  \verb|                 int nf, int nloops);|
\item
  \verb|double DecMqDownOS(double mq, double asmu, double Mth, double muth,|
  \\
  \verb|                   int nf, int nloops);|
\item
  \verb|double DecMqUpOS(double mq, double asmu, double Mth, double muth,|
  \\
  \verb|                 int nf, int nloops);|
\end{itemize}

In {\tt CRunDec} the functions \verb|DecAsDownOS| and \verb|DecAsUpOS| 
also contain the four-loop decoupling relations which have been
computed in Refs.~\cite{Schroder:2005hy,Chetyrkin:2005ia}.
Note that in the functions of this subsection the parameters $n_f$ refers to
the number of flavours in the effective theory~\cite{Chetyrkin:2000yt}.

\subsection{Functions related to the combination of running and decoupling}

\begin{itemize}
\item
  \verb|double AlL2AlH(double asmu0, double mu0, TriplenfMmu decpar[],|
  \\
  \verb|               double mu1, int nloops);|
\item
  \verb|double AlH2AlL(double asmu0, double mu0, TriplenfMmu decpar[],|
  \\
  \verb|               double mu1, int nloops);|
\item
  \verb|double mL2mH(double mq0, double asmu0, double mu0,|
  \\
  \verb|              TriplenfMmu decpar[], double mu1, int nloops);|
\item
  \verb|double mH2mL(double mq0, double asmu0, double mu0,|
  \\
  \verb|              TriplenfMmu decpar[], double mu1, int nloops);|
\end{itemize}

The parameters governing the decoupling are contained in the array
\verb|decpar[]| where each element contains the triple
$\{n_f, M_{\rm th}, \mu_{\rm th}\}$ which is realized in the structure
\begin{verbatim}
struct TriplenfMmu {
      int nf;
      double Mth;
      double muth;
};
\end{verbatim}
There is a pre-defined variable \verb|TriplenfMmu nfMmu[4];|
which can be used when calling the above functions.
Note that the components of \verb|decpar| are set to zero at the end
of the above functions.

In {\tt CRunDec} we refrain to implement the function
\verb|AsRunDec| which automatically determines the number of
active flavours for the initial and final energy scale
and performs the corresponding running and decoupling steps. In
practice it turns out 
that the decoupling of the heavy quark with mass $M_{\rm th}$ at the scale
$\mu_{\rm th}=M_{\rm th}$ is not convenient for all applications. 
Furthermore, there are
ambiguities as far as the number of active flavours is concerned in
case $\alpha_s(M_{\rm th})$ has to be evaluated using \verb|AsRunDec|.
Thus, it is recommended to use \verb|AlL2AlH| and \verb|AlH2AlL|
instead.


\section{\label{sec::examples}Typical examples}

In this section we present some typical examples which exemplify the
usage of {\tt CRunDec}. In the following we 
only display the part of
the code related to {\tt CRunDec}; the complete
programs can be found in the file {\tt example.cc} which comes
together with {\tt CRunDec}.

\subsection*{Running of $\alpha_s$ from $M_t$ to $M_b$ with five active flavours}

It is either possible to create a {\tt CRunDec} object where five 
active flavours are already specified
\begin{verbatim}
CRunDec* pObjnf5 = new CRunDec(5);
\end{verbatim}
or leave $n_f$ unspecified
\begin{verbatim}
CRunDec* pObj = new CRunDec();
\end{verbatim}
In the former case $\alpha_s^{(5)}(M_b)$ is computed from
$\alpha_s^{(5)}(M_t)=0.108$ via
\begin{verbatim}
pObjnf5 -> AlphasExact(0.108,Mz,Mb,4);
\end{verbatim}
whereas in the latter case one has
\begin{verbatim}
pObj -> AlphasExact(0.108,Mz,Mb,5,4);
\end{verbatim}
In both evaluations four-loop accuracy is assumed leading to the
result $\alpha_s^{(4)}(M_b) = 0.183$.

The number of active flavours for the object \verb|pObjnf5| can be
obtained with the help of \verb|pObjnf5 -> GetNf();| and 
\verb|pObj -> SetNf(5);| sets $n_f=5$ for \verb|pObj|.

\subsection*{Compute $\alpha_s^{(5)}(M_Z)$ from $\alpha_s^{(3)}(M_\tau)$}

Assuming a value of the strong coupling as extracted form $\tau$ decay
as $\alpha_s^{(3)}(M_\tau)=0.332$~\cite{Baikov:2008jh} 
the task is the computation of
$\alpha_s^{(5)}(M_Z)$. If the decoupling of the charm and bottom quark
is performed for $\mu_{\rm th}=2M_c$ and $\mu_{\rm th}=M_b$,
respectively, one has to specify
\begin{verbatim}
CRunDec crundec;
crundec.nfMmu[0].nf   = 4;
crundec.nfMmu[0].Mth  = Mc;
crundec.nfMmu[0].muth = 2*Mc;
crundec.nfMmu[1].nf   = 5;
crundec.nfMmu[1].Mth  = Mb;
crundec.nfMmu[1].muth = Mb;
\end{verbatim}
Afterwards $\alpha_s^{(5)}(M_Z)$ is computed via
\begin{verbatim}
crundec.AlL2AlH(0.332, Mtau,crundec.nfMmu,Mz,4);
\end{verbatim}
where four-loop accuracy for the running (corresponding to three-loop
decoupling relations) has been assumed.
As a result one obtains $\alpha_s^{(5)}(M_Z)=0.1200$.

\subsection*{Compute $m_b(m_b)$ from $m_b(10~\mbox{GeV})$}

In Ref.~\cite{Chetyrkin:2009fv} the $\overline{\rm MS}$ bottom quark mass 
has been extracted for $\mu=10$~GeV as $m_b(10~\mbox{GeV})=3.610\pm 0.016$~GeV.
The scale-invariant mass $m_b(m_b)$ can been computed with four-loop
accuracy via
\begin{verbatim}
CRunDec cmpasmb(5);
double asmu = cmpasmb.AlphasExact(0.1189,Mz,10,4);
cmpasmb.mMS2mSI(3.610,asmu,10,5,4);
\end{verbatim}
where $\alpha_s^{(5)}(M_Z)=0.1189$ has been used.
The result reads $m_b(m_b) = 4.163$~GeV.
In the {\tt Mathematica} version of {\tt RunDec} it is not recommended
to use \verb|mMS2mSI| since in intermediate steps $\Lambda_{\rm QCD}$ 
is used and thus the final result does not have the required precision.

\subsection*{Compute $m_t(m_t)$ from the top quark on-shell mass}

The on-shell top quark mass has been measured at the {\tt Tevatron} 
experiments {\tt D0} and {\tt CDF} to
$M_t=173.2\pm0.9$~GeV~\cite{Lancaster:2011wr}. If this value shall be
transformed to $m_t(m_t)$ one proceeds in the following way
\begin{verbatim}
CRunDec cmpmt(6);
double  MtOS    = 173.2;

cmpmt.nfMmu[0].nf   = 6;
cmpmt.nfMmu[0].Mth  = MtOS;
cmpmt.nfMmu[0].muth = MtOS;

double as6Mt = cmpmt.AlL2AlH(asMz,Mz,cmpmt.nfMmu,MtOS,4);
double mtMt  = cmpmt.mOS2mMS(MtOS,cmpmt.mq,as6Mt,MtOS,3);
double mtmt  = cmpmt.mMS2mSI(mtMt,as6Mt,MtOS,4);
\end{verbatim}
The final result reads $m_t(m_t)=164.0$~GeV.

\subsection*{Compute on-shell charm and bottom quark mass from $m_c(m_c)$ and
  $m_b(m_b)$}

After defining a pointer (\verb|pObjnf5|) to a \verb|CRunDec| object with five
active flavours (as in the example above) one obtains 
$\alpha_s^{(5)}(m_b(m_b))$ via
\begin{verbatim}
double alpha5mub = pObjnf5 -> AlphasExact(asMz, Mz, mub, 4);
\end{verbatim}
and subsequently the on-shell mass $M_b$ with two and three-loop accuracy
with the help of
\begin{verbatim}
double mbOS2 = pObjnf5 -> mMS2mOS(mub, pObjnf5->mq, alpha5mub, mub, 2);
double mbOS3 = pObjnf5 -> mMS2mOS(mub, pObjnf5->mq, alpha5mub, mub, 3);
\end{verbatim}
The results read $M_b = 4.762$~GeV and $M_b = 4.909$~GeV, respectively.

In the case of the charm quark one proceeds in analogy by evaluating in a
first step $\alpha_s^{(4)}(m_c(m_c))$ 
\begin{verbatim}
pObjnf5 -> nfMmu[0].nf   = 4;
pObjnf5 -> nfMmu[0].Mth  = Mc;
pObjnf5 -> nfMmu[0].muth = 2*Mc;
double alpha4muc = pObjnf5 -> AlH2AlL(asMz, Mz, pObjnf5 -> nfMmu, mub, 4);
\end{verbatim}
and afterwards the on-shell mass $M_c$ to two and three loops
\begin{verbatim}
double mcOS2 = pObjnf5 -> mMS2mOS(muc, pObjnf5->mq, alpha4muc, muc, 2);
double mcOS3 = pObjnf5 -> mMS2mOS(muc, pObjnf5->mq, alpha4muc, muc, 3);
\end{verbatim}
leading to $M_c = 1.494$~GeV and $M_c = 1.573$~GeV.



\section*{Acknowledgements}

We thank Konstantin Chetyrkin and Johann K\"uhn for discussions and useful
comments. This work was supported by the BMBF through Grant No. 05H09VKE.




\end{document}